\documentclass[intlimits,twoside,a4paper]{article}
\usepackage{amssymb,amsmath}
\usepackage{wrapfig}
\usepackage{graphicx}

\usepackage[T2A]{fontenc}
\usepackage[cp1251]{inputenc}

\usepackage{cmpj2}

\newcommand{\be}{\begin{equation}}
\newcommand{\ee}{\end{equation}}
\newcommand{\bea}{\begin{eqnarray}}
\newcommand{\eea}{\end{eqnarray}}
\newcommand{\non}{\nonumber\\}

\issue{2014}{17}{4}{43703}
\doinumber{10.5488/CMP.17.43703}

\title[Electrocaloric effect in KH$_2$PO$_4$ family crystals]{Electrocaloric effect in KH$_2$PO$_4$ family crystals}
\author[A.S. Vdovych \textsl{et al.}]{A.S. Vdovych\refaddr{label1}, A.P. Moina\refaddr{label1}, R.R. Levitskii\refaddr{label1},
        I.R. Zachek\refaddr{label2}}
\addresses{
\addr{label1} Institute for Condensed Matter Physics of the
National Academy of Sciences of Ukraine, 1 Svientsitskii~St.,
79011 Lviv, Ukraine
\addr{label2} National University ``Lviv Polytechnic'', 12 Bandera St., 79013 Lviv, Ukraine }

\date{Received May 31, 2014, in final form October 7, 2014}
\authorcopyright{A.S. Vdovych, A.P. Moina, R.R. Levitskii, I.R. Zachek, 2014}
\begin{document}

\maketitle

\begin{abstract}
The proton ordering model for the KH$_{2}$PO$_{4}$ type
ferroelectrics is modified by taking into account the dependence of the effective dipole moments on
the proton ordering parameter. Within the four-particle cluster
approximation we calculate the crystal polarization and explore
the electrocaloric effect. Smearing of the ferroelectric phase
transition by a longitudinal electric field is described. A good
agreement with experiment is obtained.
\pacs 77.84.Fa, 77.70.+a

\end{abstract}

\section{Introduction}

The electrocaloric (EC) effect is the change of temperature of a
dielectric at an adiabatic change of the applied electric field.
Research in this field is driven by a quest for materials that can
be used for efficient, environment-friendly, and compact (on-chip)
solid-state cooling devices.

 The current state of the art on the electrocaloric effect
research for ferroelectrics is well summarized in
\cite{Scott,Valant}. At the moment, the largest effect is observed
in perovskite ferroelectrics. Thus, in \cite{Mischenko2006} in the
PbZr$_{0.95}$Ti$_{0.05}$O$_{3}$ thin film with a thickness of
350~nm in a strong electric field (480~kV/cm) the obtained
electrocaloric temperature change is $\Delta T=12$~K. Ab
initio molecular dynamics calculations \cite{Rose2012} predict
$\Delta T \thickapprox 20$~K in LiNbO$_3$. In the hydrogen bonded
ferroelectrics of the KH$_{2}$PO$_{4}$ (KDP) type, the
electrocaloric effect was studied for relatively low fields only.
Thus, it has been obtained that $\Delta T \thickapprox0.04$~K at
$E\thickapprox 4$~kV/cm \cite{363x}, $\Delta T \thickapprox~1$~K
at $E\thickapprox 12$~kV/cm \cite{Baumgartner1950}, and $\Delta T
\thickapprox0.25$~K at $T_\textrm{c}$ and $E\thickapprox 1.2$~kV/cm
\cite{Shimshoni1969}.

Theoretical calculations of the electrocaloric effect in KDP have
been made in  \cite{Dunne2008} within the Slater model
\cite{Slater1941} and in  the paraelectric phase only. It is also
known that the Slater model gives incorrect results in the
ferroelectric phase, and more complicated versions of the proton
ordering model are required for an adequate description of these
crystals. Thus, the effect of electric field on the physical
characteristics of the KDP type crystals, such as polarization,
dielectric permittivity, piezoelectric coefficients, elastic
constants,  has been described within the proton ordering model
with the piezoelectric coupling to the shear strain
  $\varepsilon_6$ \cite{Stasyuk2001,0311U6,JPS1701} and with proton tunneling \cite{lis2007} taken into account.
   However, these theories required, in particular, invoking two different values
  of   the effective dipole moments for the paraelectric and ferroelectric phase \cite{Stasyuk2001,JPS1701}.
  This made impossible a correct description of the system
  behavior in the fields high enough to smear out the first order phase transition.
  There is an inner logical contradiction in the model: while  no physical characteristic of a
crystal should exhibit any discontinuity in the fields above the
critical one, there is no smooth transition between the values of
model parameters, rigidly set to be different for the two phases.

In the present paper we suggest a way to remove this
contradiction. Assuming that the difference between the dipole
moments is caused by non-zero values of the order parameter, we
modify the proton ordering model accordingly.
 The field dependences of polarization, smearing of the first order phase transition, and the electrocaloric effect are described.

\clearpage

\section{Thermodynamic characteristics}

We consider the KDP type ferroelectrics in the presence of an
external electric field $E_3$ applied along the crystallographic
axis
 $\textbf{c}$, inducing the strain
 $\varepsilon_6$ and polarization $P_3$.
The total model Hamiltonian reads
 \be
 \label{ham0}
 \hat H = N\hat H_{0} + \hat H_\textrm{s}\,,
 \ee
 where $N$ is the total number of primitive cells.  The ``seed''
energy  $H_{0}$ corresponds to the sublattice of heavy ions and
does not explicitly depend on the proton subsystem configuration.
It is expressed in terms of the strain $\varepsilon_6$ and
electric field
 $E_3$ and includes the elastic, piezoelectric, and dielectric
 contributions \cite{0311U6}
 \be
 \hat H_{0} = {v} \left( \frac12 c_{66}^{E0}\varepsilon_6^2 -
 e_{36}^0E_3\varepsilon_6 - \frac12 \chi_{33}^{\varepsilon 0}E_3^2
 \right),
 \ee
 where $v$ is the
primitive cell volume; $c_{44}^{E0} $, $e_{36}^0$, ${\chi }_{33}^{\varepsilon 0} $ are the ``seed'' elastic constant,
piezoelectric coefficient, and dielectric susceptibility, respectively.

The pseudospin part of the Hamiltonian reads
 \be
 \hat H_\textrm{s} = \frac12 \sum\limits_{ {qf},{q'f'}}J_{ff'}(qq')
 \frac{\sigma_{qf}}{2}\frac{\sigma_{q'f'}}{2} + \hat H_\textrm{sh} + \sum\limits_{qf} 2\psi_6\varepsilon_6 \frac{\sigma_{qf}}{2}
 - \sum\limits_{qf}\mu_{f}E_3 \frac{\sigma_{qf}}{2} + \hat H_E\,.
 \label{H_s}
 \ee
Here, the first term describes the effective long-range
interactions between  protons, including also indirect
lattice-mediated interactions  \cite{122x,133x}; $\sigma_{qf}$ is
the operator of the
 $z$-component of a pseudospin, corresponding to the proton on the $f$-th hydrogen bond ($f=1,\,2,\,3,\,4$) in the  $q$-th
cell. Its eigenvalues $\sigma_{qf}= \pm 1$
  are assigned to two equilibrium positions of a proton on this bond.

In (\ref{H_s}), $\hat H_\textrm{sh}$ is the Hamiltonian of short-range interactions between
 protons, which includes  terms  linear over the strain
\cite{0311U6}
 \bea
 \hat H_\textrm{sh} &=& \sum\limits_q \bigg\{ \left(
 \frac{\delta_{s}}{8}\varepsilon_6 +
 \frac{\delta_{1}}{4}\varepsilon_6\right) (\sigma_{q1} + \sigma_{q2} + \sigma_{q3} +
 \sigma_{q4})  \non
 && +\left( \frac{\delta_{s}}{8}\varepsilon_6 - \frac{\delta_{1}}{4}\varepsilon_6\right)
 (\sigma_{q1} \sigma_{q2} \sigma_{q3} + \sigma_{q1} \sigma_{q2} \sigma_{q4}
 + \sigma_{q1} \sigma_{q3} \sigma_{q4} + \sigma_{q2} \sigma_{q3}
 \sigma_{q4})  \non
 &&  + \frac14 (V + \delta_{a}\varepsilon_6)(\sigma_{q1} \sigma_{q2} + \sigma_{q3} \sigma_{q4}
 )+ \frac14 (V - \delta_{a}\varepsilon_6)(\sigma_{q2} \sigma_{q3} + \sigma_{q4}
 \sigma_{q1}) \nonumber \\
 &&  + \frac U4  (\sigma_{q1} \sigma_{q3} + \sigma_{q2}
 \sigma_{q4}) + \frac{\Phi}{16}  \sigma_{q1} \sigma_{q2} \sigma_{q3}
 \sigma_{q4}\bigg\}.
 \eea
Here,
 \[
 V = - \frac12 w_1\,, \qquad U = \frac12 w_1 - \varepsilon\,, \qquad \Phi =
 4\varepsilon- 8w + 2w_1\,,
 \]
and $\varepsilon$, $w$, $w_1$ are the energies of proton
configurations.

The third term in (\ref{H_s}) is a linear over the shear strain
$\varepsilon_6$ field due to the piezoelectric coupling; $\psi_6$
is the deformational potential. The fourth term effectively
describes the system interaction with the external electric field
$E_3$. Here, $\mu_{f}$ is the effective dipole moment of the
$f$-the hydrogen bond, and
 \[
 \mu_{1} = \mu_{2} = \mu_{3} = \mu_{4} = \mu.
 \]

The fifth term in (\ref{H_s}) is introduced in the present paper
for the first time. It takes into account the assumed dependence
of the effective dipole moment on the order parameter (pseudospin
mean value)
\be \hat H_E =-
\bigg(\frac{1}{N}\sum\limits_{q'f'}\frac{\sigma_{q'f'}}{2}\bigg)^2
\mu'E_3\sum\limits_{qf} \frac{\sigma_{qf}}{2}\,. \label{H_E}\ee
It is equivalent to a term proportional to $P_3^3E_3$ in a
phenomenological thermodynamic potential. Note that the terms like
$P_3^2E_3$ are not allowed because of the symmetry considerations,
and we keep the Hamiltonian to be linear in the field $E_3$.

In view of the crystal structure of the KDP type ferroelectrics,
the four-particle cluster approximation is most suitable for
short-range interactions \cite{133x,46x}. Long-range
interactions and the term $\hat H_E$ are taken into account in the
mean field approximation. Thus,
 \be
 \label{munew}
 \hat H_E
 \approx
  - 12N\mu'E_3 \eta^2\sum\limits_{f=1}^4 \frac{\sigma_{qf}}{2}
+ 16N\mu'E_3 \eta^3.
 \ee
Combining the fourth term in (\ref{H_s}) and the first term  in
(\ref{munew}), we obtain the following term in the Hamiltonian $-
(\mu+12\mu'\eta^2)E_3\sum_{qf} {\sigma_{qf}}/{2}$.
Effectively, the term $12\mu'\eta^2$  in $(\mu+12\mu'\eta^2)$
describes the jump of the dipole moment at the first order phase
transition, its different values for the paraelectric and
ferroelectric phase, and its smooth behavior in the fields above
the critical one, when there is no jump of $\eta$. We can now use
a single value of $\mu$ for both phases and remove the logical
contradiction of the earlier theories, described in Introduction.

Proceeding with the standard  calculations of the  cluster
approximation \cite{Stasyuk2001,JPS1701,46x}, we obtain the
following expression for the proton ordering parameter
\[
 \eta = \langle \sigma_{q1} \rangle = \langle \sigma_{q2} \rangle
 = \langle \sigma_{q3} \rangle = \langle \sigma_{q4} \rangle =\frac{m
 }{D}\,,
 \]
where
 \begin{align}
 m  &= \sinh (2z  + \beta \delta_{s}\varepsilon_6) + 2b \sinh(z  - \beta
 \delta_{1}\varepsilon_6), \non[1ex]
 D &= \cosh (2z  + \beta \delta_{s}\varepsilon_6) + 4b \cosh
 (z  - \beta \delta_{1}\varepsilon_6) + 2a \cosh \beta
 \delta_{a}\varepsilon_6 + d, \non
  z  &= \frac12 \ln \frac{1 + \eta}{1 - \eta}
 + \beta \nu_\textrm{c}\eta - \beta\psi_6\varepsilon_6 + \frac{\beta
 \mu }{2}E_3  + 6\beta\mu'\eta^2 E_3\,, \non
 a &= \re^{-\beta \varepsilon}, \qquad b = \re^{-\beta w}, \qquad d = \re^{-\beta
 w_1}; \nonumber
 \end{align}
 $4\nu_\textrm{c} =
J_{11}(0) + 2J_{12}(0) + J_{13}(0)$ is the eigenvalue  of the
long-range interactions matrix Fourier transform  $J_{ff'} =
\sum_{ {\bf R}_q - {\bf R}_{q'}} J_{ff'}(qq')$;   $\beta={1}/{k_\textrm{B}T}$.

The thermodynamic potential  is then obtained in the following
form
 \bea
  G &=& \frac{ {v}}{2}c^{E0}_{66}\varepsilon^2_6 -
{{v}}e^0_{36}\varepsilon_6E_3
- \frac{{v}}{2}\chi^{\varepsilon 0}_{33}E^2_3 + 2{{\nu}}_\textrm{c}\eta^2 + 16\mu'E_3 \eta^3 \label{G}\\
&&+ \frac2\beta\ln 2 - \frac2\beta\ln\left(1- \eta^2\right) -
\frac2\beta\ln D- {{v}}\sigma_6\varepsilon_6\,. \nonumber
\eea
Here, $\sigma_6$ is the formally introduced shear stress conjugate to
the strain $\varepsilon_6$. In numerical calculations we put
$\sigma_6=0$. The condition of the thermodynamic potential minimum
\[
 \left( \frac{\partial
G}{\partial\varepsilon_6} \right)_{T,E_{3},\sigma_6} = 0
\]
yields an equation for the strain  $\varepsilon_6$
 \bea
&& \hspace{-8ex} \sigma_6 = c^{E0}_{66}\varepsilon_6  -
e^0_{36}E_3  +  \frac{4\psi_6}{v} \eta  + \frac{2r }{ v D}\,.
\label{s_6}
 \eea
In the same way, we derive the expressions for polarization $P_3$
and molar entropy of the proton subsystem
\bea
   P_3 &=& - \frac{1}{{v}}\left( \frac{\partial G}{\partial E_3}
\right)_{T,\sigma_6} = e^0_{36}\varepsilon_6 + \chi^{\varepsilon 0}_{33}E_3 +
2\frac{\mu}{v}\eta +
8\frac{\mu'}{v}\eta^3, \label{P_3} \\
 S  &=& - \frac{N_\textrm{A}}2\left( \frac{\partial
G}{\partial T}\right)_{E_3,\sigma_6} = R \left[ -\ln2 + \ln (1 -
\eta^2) + \ln D +  2Tz_T\eta + \frac{M }{D} \right]. \label{S}
 \eea
Here, $N_\textrm{A}$ is the Avogadro number; $R$ is the gas constant. The
following notations are used:
\begin{align}
 r &=-\delta_{s}M_{s}-\delta_{a}M_{a}+\delta_{1}M_{1}\,, \nonumber \\
z_T &= - \frac{1}{k_\textrm{B} T^2}({\nu}_\textrm{c} \eta -
\psi_6\varepsilon_6 + 6\mu'\eta^2 E_3), \nonumber
\end{align}
\begin{align}
 M  &= 4b\beta w\cosh(z  - \beta
 \delta_{1}\varepsilon_6) + \beta w_1d + 2a\beta\varepsilon\cosh\beta
 \delta_{a}\varepsilon_6+\beta
 \varepsilon_6r , \nonumber \\
M_{a}&=2a\sinh \beta\delta_{a}\varepsilon_6, M_{s}=\sinh (2z
+\beta\delta_{s}\varepsilon_6), M_{1}=4b \sinh (z
-\beta\delta_{1}\varepsilon_6). \nonumber
\end{align}

Expressions for  dielectric susceptibilities, piezoelectric
coefficients, and elastic constants derived  \cite{ec-preprint}
 from equations (\ref{s_6}), (\ref{P_3}) are slightly different from the
previous ones \cite{Stasyuk2001}, where the dependence of the
effective dipole moment on the order parameter was not taken into
account. Numerical calculations, however, showed
\cite{ec-preprint} that in zero electric field the difference is
minor.

The molar specific heat of the subsystem described by the
Hamiltonian (\ref{ham0}) is
 \be
\label{csigma}
 \Delta
C^\sigma = T\left( \frac{\partial S}{\partial T}\right)_{\sigma} =
T(S_T + S_{\eta}\eta_T + S_{\varepsilon}\varepsilon_T). \ee
Here,
\bea && S_T = \left(\frac{\partial S}{\partial T}
\right)_{P_3,\varepsilon_6} =\frac{R}{DT}\left[ 2Tz_T (q_6 - \eta
M ) +
N_6 - \frac{M ^2}{D} \right], \nonumber\\
&& S_{\eta} = \left(\frac{\partial S}{\partial \eta}
\right)_{\varepsilon_6,T} = \frac{2R}{D} \left[ DTz_T + (q_6 - \eta M
) z_{\eta}\right] ,\nonumber \\
&& \label{csigma1}
 S_{\varepsilon}  =  \left(\frac{\partial S}{\partial\varepsilon_6}\right)_{\eta,T}  =
\frac{R}{DT}\left[ - 2\left(q_6 - \eta M \right) \psi_6 -  \lambda
 + \frac{M }{D} r  \right].
\end{eqnarray}
Notations introduced here are described in appendix.

Then, the total specific heat is
\be C=\Delta C^\sigma + C_\textrm{regular}\,.
\ee
Here, $\Delta C^\sigma$
is assumed to describe all the anomalies of the specific heat at
the phase transition, whereas the regular background contribution
to the specific heat, mostly from the lattice of heavy ions, is
approximated by a linear temperature dependence \be \label{creg}
C_\textrm{regular} = C_0+C_1(T-T_\textrm{c}). \ee As will be discussed later, this
linear approximation agrees with the experimental data.

Finally, the electrocaloric temperature change is calculated using
the known formula
\be \Delta T = \int \limits_{0}^{E_3} \frac{TV}{C}
\left(\frac{\partial P_3}{\partial T} \right)_{E}\rd E_3\,,
\label{DT_int}\ee
where the pyroelectric coefficient is
\be \left(\frac{\partial P_3}{\partial T} \right)_{E} =
e_{36}^{0}\varepsilon_T + \frac{2(\mu+12\mu'\eta^2)}{v}\eta_T\,, \ee
$V=vN_\textrm{A}/2$ is the molar volume.

\section{Numerical calculations}

To perform the numerical calculations we need to set the values of the following theory parameters:
 \begin{enumerate}
 \item[---] the Slater energies
 $\varepsilon$, $w$, $w_{1}$;
 \item[---] the parameter of the long-range interactions $\nu_\textrm{c}$;
 \item[---] the effective dipole moment $\mu$ and  the correction is due to proton ordering $\mu'$;
 \item[---] the deformation potentials  $\psi_{6}$,
 $\delta_{s}$, $\delta_{a}$,
$\delta_{1}$;
 \item[---] the ``seed''  dielectric susceptibility
 $\chi_{33}^{\varepsilon 0}$, elastic constant $c_{66}^{E0}$, piezoelectric coefficient
 $e_{36}^0$;
  \item[---] the parameters of the lattice specific heat $C_0$ and $C_1$.
 \end{enumerate}
They are chosen, obviously, by fitting the theoretical
thermodynamic characteristics to the experimental data, as
described in  \cite{JPS1701}. The obtained optimum sets of the
model parameters are given in table~\ref{tab1}.

To describe crystals with different deuteration levels, we use the
mean crystal approximation, where the theory parameters are
assumed to be linearly dependent on deuteron concentration (except
for the parameter $\nu_\textrm{c}$, for which a small deviation from the
linear dependence is assumed, as it is chosen from the
condition that the calculated transition temperature coincides
with the experimental one, which is also slightly non-linear). The
dependence of the energy levels and interparticle interaction
constants on deuteration is caused by the corresponding
geometrical changes in the crystal structure with deuteration
(elongation of the hydrogen bonds, changes in the distance between
the equilibrium positions of H or D on the bonds, changes in the
lattice constants, etc).

\begin{table}[!h]
\caption{The optimum sets of the model parameters for different
crystals. As KD$_{2}$PO$_{4}$ we denoted
K(H$_{1-x}$D$_{x})_{2}$PO$_{4}$ with $x=0.89$.\label{tab1}}
\vspace{2ex}
\begin{center}
\begin{tabular}{c|c|c|c|c|c|c|c}
\hline\hline  &  $T_\textrm{c}^0$ & ${\varepsilon}/{k_\textrm{B}}$ & ${w}/{k_\textrm{B}}$ & ${\nu_\textrm{c}}/{k_\textrm{B}}$ & $\mu$ & $\mu'$ & $\chi_{33}^0$ \\
 & (K) & (K) & (K) & (K) & ($10^{-30}$~C$\cdot $m) & ($10^{-30}$~C$\cdot $m) & \\
\hline\hline KH$_{2}$PO$_{4}$  & 122.22 & 56.00 & 430.0 & 17.55 & 5.6 &  $-0.217$ &  0.75 \\
KD$_{2}$PO$_{4}$  & 211.73 & 85.33 & 730.4 & 39.26 & 6.8 &  $-0.217$ &  0.39 \\
\hline KH$_{2}$AsO$_{4}$    & 97 & 35.50 & 385.0 & 17.43 & 5.5 &  $-0.033$ &  0.7 \\
       KD$_{2}$AsO$_{4}$   & 162 & 56.00 & 690.0 & 31.72 & 7.3 &  $-0.000$ &  0.5 \\
 \hline\hline
\end{tabular}
\end{center}
\begin{center}
\begin{tabular}{c|c|c|c|c|c|c|c|c}
\hline\hline &  ${\psi_{6}}/{k_\textrm{B}}$ & ${\delta_{s}}/{k_\textrm{B}}$ & ${\delta_{a}}/{k_\textrm{B}}$ & ${\delta_{1}}/{k_\textrm{B}}$ & $c_{66}^{E0}$ & $e_{36}^0$ &$C_0$ &$C_1$ \\
  & (K) & (K) & (K) & (K) & ($10^{9}$ N/m$^2$) & (C/m$^2$) & J/(mol K) & J/(mol K$^2$) \\
 \hline\hline KH$_{2}$PO$_{4}$ &$-150.00$ & 82.00 &$-500.00$ &$-400.0$ &  7.00 &0.0033 & 60 & 0.32\\
 KD$_{2}$PO$_{4}$ &$-139.89$ & 48.64 &$-1005.68$ &$-400.0$ &  6.39 &0.0033 & 93 & 0.32 \\
 \hline KH$_{2}$AsO$_{4}$   &$-170.00$ & 130.00 &$-500.0$ &$-500.0$ &  7.50 &0.01 & 60 &0.32 \\
        KD$_{2}$AsO$_{4}$   &$-160.00$ & 120.00 &$-800.0$ &$-500.0$ &  6.95 &0.01 & 98 & 0.40\\
 \hline\hline
\end{tabular}
\end{center}
\end{table}

The primitive cell volume is taken to be $v=0.1946\cdot 10^{-21}$
cm$^3$ for K(H$_{1 - x}$D$_{x})_{2}$PO$_{4}$ and $v=0.202\cdot
10^{-21}$~cm$^3$ for K(H$_{1 - x}$D$_{x})_{2}$AsO$_{4}$,
irrespectively of the deuteration. The energy $w_{1}$ of proton
configurations with four or zero protons near the given oxygen
tetrahedron should be much higher than $\varepsilon$ and $w$.
Therefore, we take $w_{1} = \infty$ $(d=0)$.


As we have already mentioned, when the dependence of the effective
dipole moment on the order parameter is taken into account, the
agreement between the theory and experiment for most of the
calculated dielectric, piezoelectric, elastic characteristics, and
specific heat of the studied crystals in the absence of an external
electric field is neither improved nor worsened (see
\cite{ec-preprint}). However, the present model allows us to
describe more consistently the smearing of the first order phase
in high electric fields.

\begin{figure}[!t]
\begin{center}
 \includegraphics[width=0.45\textwidth]{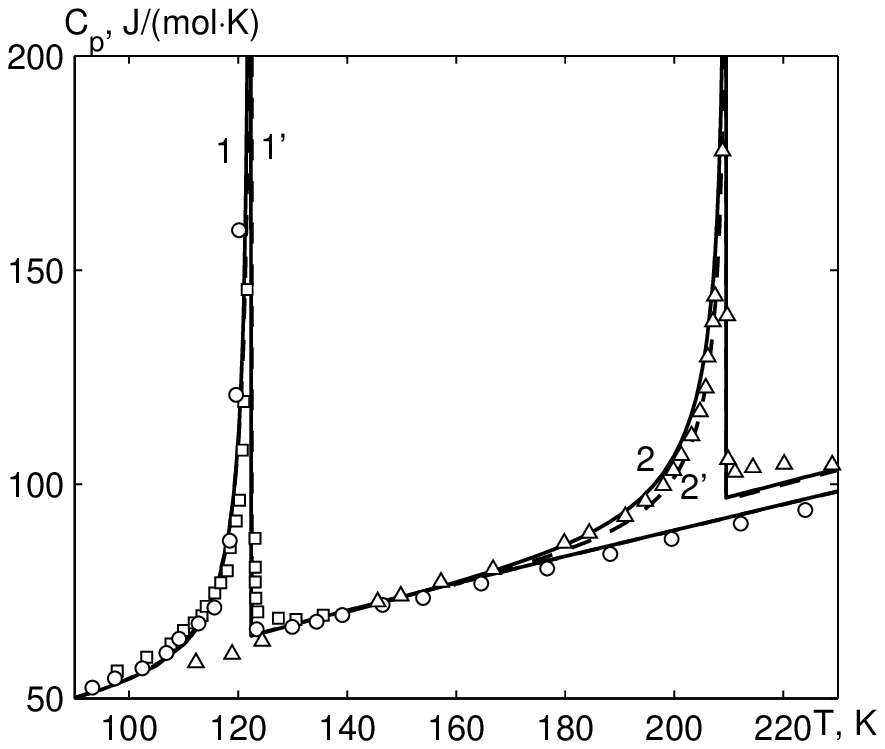}
\end{center}
\caption{The temperature dependence of the molar specific heat
of  K(H$_{1 - x}$D$_{x})_{2}$PO$_{4}$ at $x=0.0$~--- $\circ$
\cite{Stephenson1397},
 $\square$ \cite{380x}; at $x=0.86$~--- $\bigtriangleup$ \cite{380x}. Dashed lines 1' and 2': the theoretical results of \cite{JPS1701}.} \label{C_x86}
\end{figure}

The temperature dependence of the specific heat of
KH$_{2}$PO$_{4}$ and KD$_{2}$PO$_{4}$ is shown in
figure~\ref{C_x86}. The contribution $\Delta C^\sigma$ is essential
in the transition region and satisfactorily describes the
experimental anomalies. As one can see, the total specific heat
above $T_\textrm{c}$ can be well approximated by a linear temperature
dependence, thus justifying the linear dependence of
$C_\textrm{regular}$, given by equation (\ref{creg}).

\begin{figure}[!b]
\begin{center}
 \includegraphics[width=0.45\textwidth]{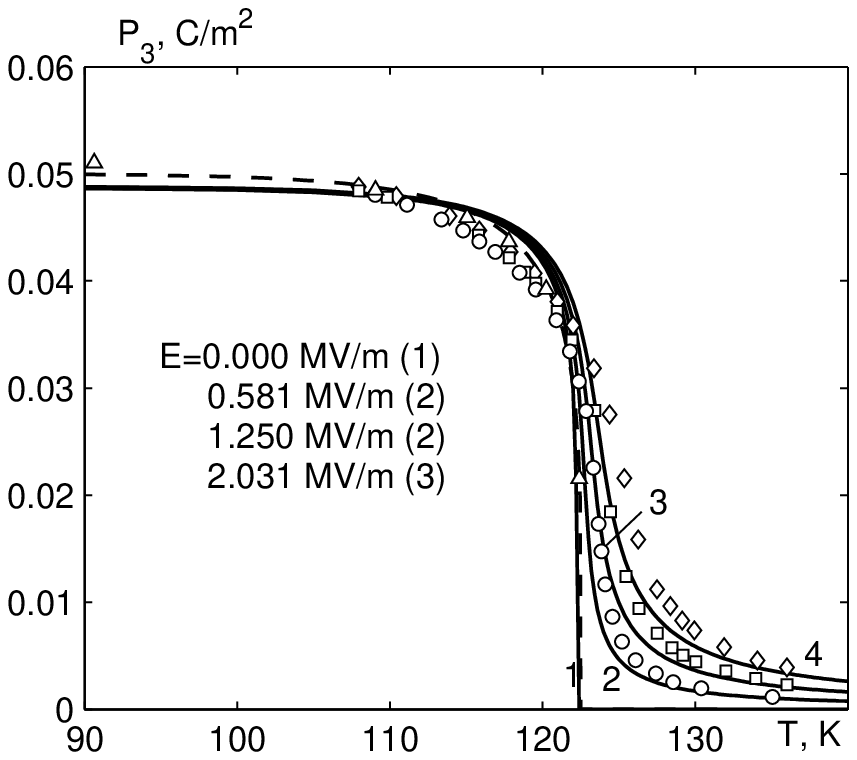}
 \hspace{5mm}
 \includegraphics[width=0.45\textwidth]{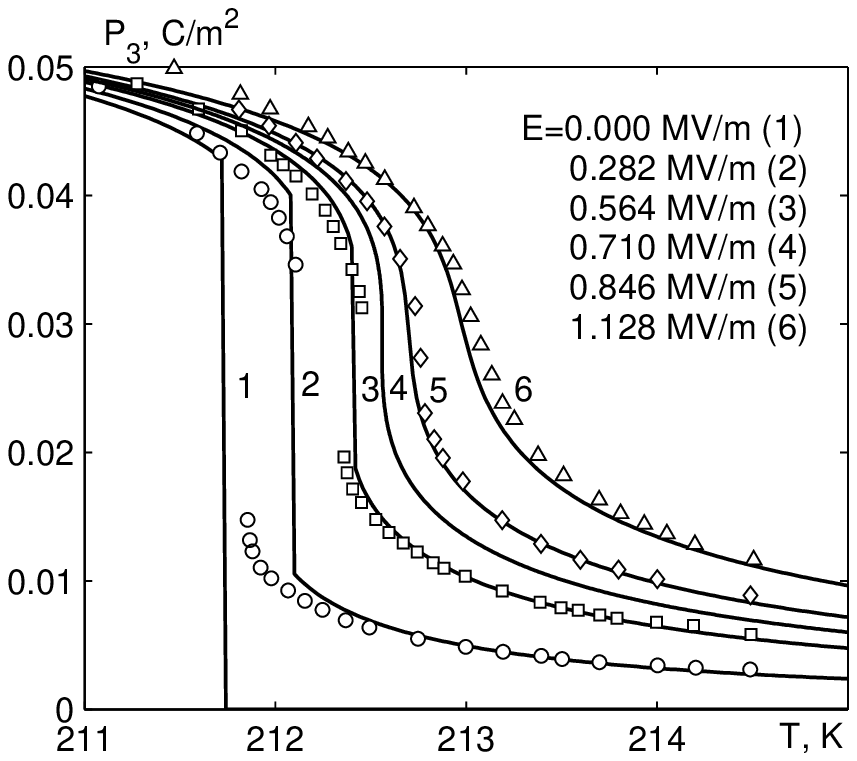}
\end{center}
\parbox[t]{0.5\textwidth}{
\caption{The temperature dependence of polarization of
KH$_{2}$PO$_{4}$  at different  $E_3$(MV/m): 0.0~--- 1,
$\vartriangle$ \cite{363x}; 0.581~--- 2, $\circ$ \cite{369x}; 1.250~--- 3, $\square$ \cite{369x}; 2.031~--- 4, $\lozenge$ \cite{369x}.
Symbols are experimental points; solid lines: the present theory;
dashed  lines: the theoretical results of  \cite{JPS1701}.} \label{Ps}
}
\parbox[t]{0.5\textwidth}{
\caption{The temperature dependence of polarization of
K(H$_{1 - x}$D$_{x})_{2}$PO$_{4}$ at $x=0.89$  and at different  $E_3$ (MV/m): 0.0~--- 1; 0.282~--- 2, $\circ$;
0.564~--- 3, $\square$; 0.71~--- 4; 0.846~--- 5, $\lozenge$; 1.128~--- 6, $\vartriangle$. Symbols are experimental points taken from \cite{Sidnenko978}; lines: the present theory.} \label{Ps_x89}
}
\end{figure}

In figures~\ref{Ps}  and \ref{Ps_x89} we plotted the temperature
variation of polarization of K(H$_{1 - x}$D$_{x})_{2}$PO$_{4}$
in different fields. The agreement with experiment is better at
$x=0.89$ (and 0.84, see \cite{ec-preprint}) than at $x=0$. We
believe this is due to proton tunnelling, essential in
non-deuterated samples, which is not included in our model.

The field $E_{3}$, which in these crystals is the field conjugate
to the order parameter, induces non-zero polarization $P_{3}$
above the transition point. Polarization has a jump at $T_\textrm{c}$,
indicating the first order phase transition. With an increasing
field, the polarization jump decreases, whereas the transition
temperature $T_\textrm{c}$ increases almost linearly. The corresponding
$\partial T_\textrm{c}/\partial E_3$ slopes are 0.192 and 0.115~K~cm/kV for
$x=0$ and $x=0.89$, respectively (c.f. 0.22 and 0.13~K~cm/kV from
our earlier calculations \cite{Stasyuk2001} and experimental
$0.125$~K~cm/kV of \cite{GladkiiSidnenko} for $x=0.89$). At some
critical field $E^*$, the jump vanishes, and the transition smears
out. The calculated coordinates of the critical point are
 $E^*=125$~V/cm, $T^*_\textrm{c}$=122.244~K for $x=0$  and $7.1$~kV/cm, 212.55~K for $x=0.89$, which agrees well with the experiment \cite{Western,GladkiiSidnenko}.  It should be noted that in our previous
calculations \cite{JPS1701} it was impossible to obtain a correct
description of the polarization behavior in the fields above the
critical one, due to the necessity of using two different values
of the effective dipole moment $\mu$ in calculations.

\begin{figure}[!t]
\begin{center}
 \includegraphics[width=0.49\textwidth]{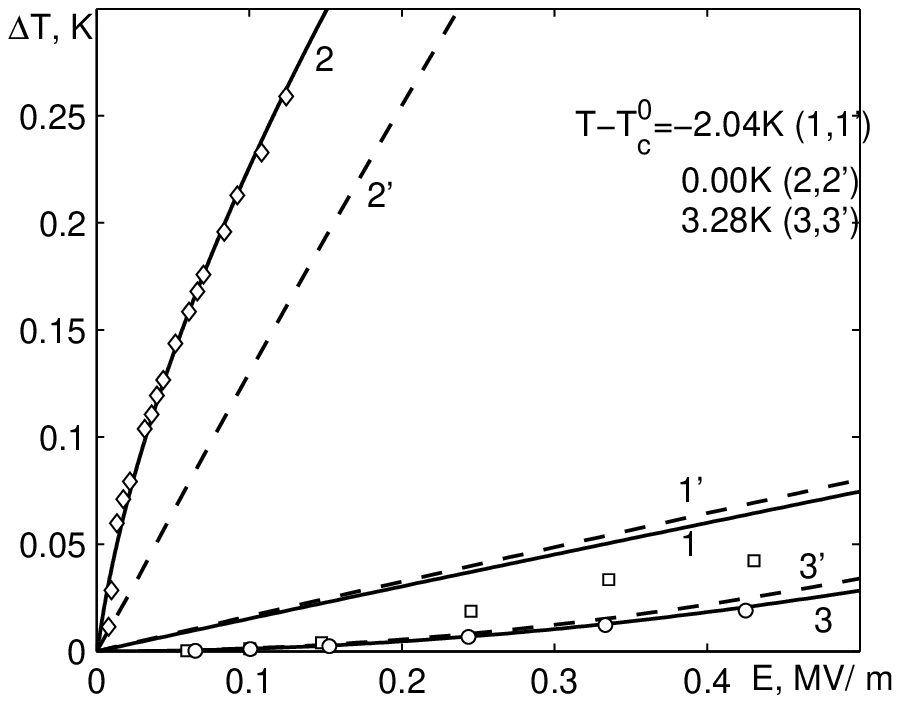}
 \hspace{4mm}
 \includegraphics[width=0.45\textwidth]{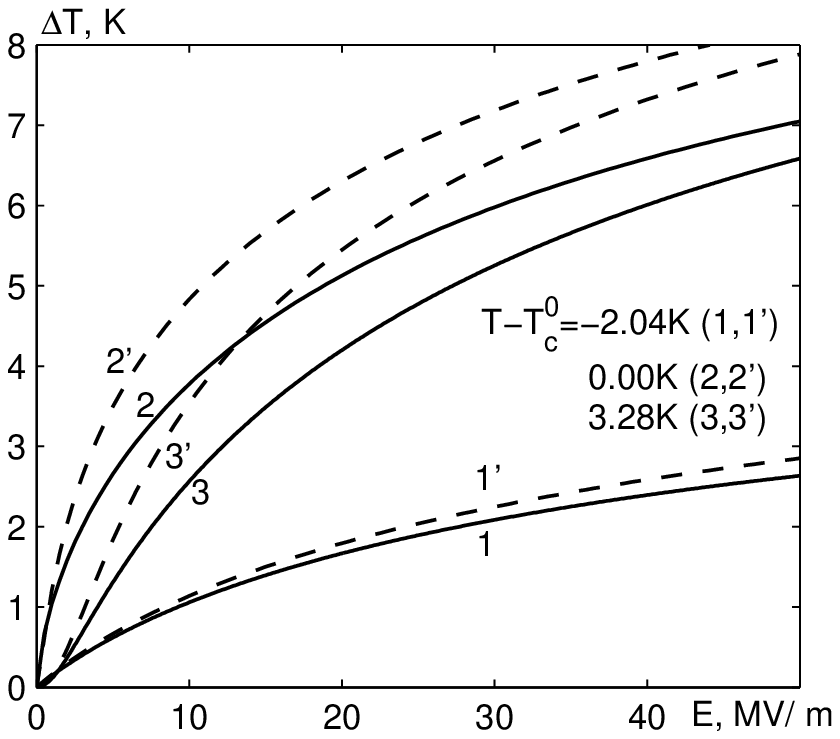}
\end{center}
\caption{The field dependence of the electrocaloric temperature
change of K(H$_{1 - x}$D$_{x})_{2}$PO$_{4}$ for $x=0.0$ (solid
lines) and  $x=0.89$ (dashed lines)  at $T-T_\textrm{c}^0=-2.04$~K~--- 1,
$1'$, $\square$; $T=T_\textrm{c}^0$~--- 2, $2'$, $\circ$; $T-T_\textrm{c}^0=3.2$~K~---
3, $3'$, $\diamond$. Experimental points are taken from
\cite{363x}~--- $\circ$, $\square$ and \cite{Shimshoni1969}~---
$\diamond$.} \label{DT_x89}
\end{figure}

\begin{figure}[!b]
\begin{center}
 \includegraphics[width=0.485\textwidth]{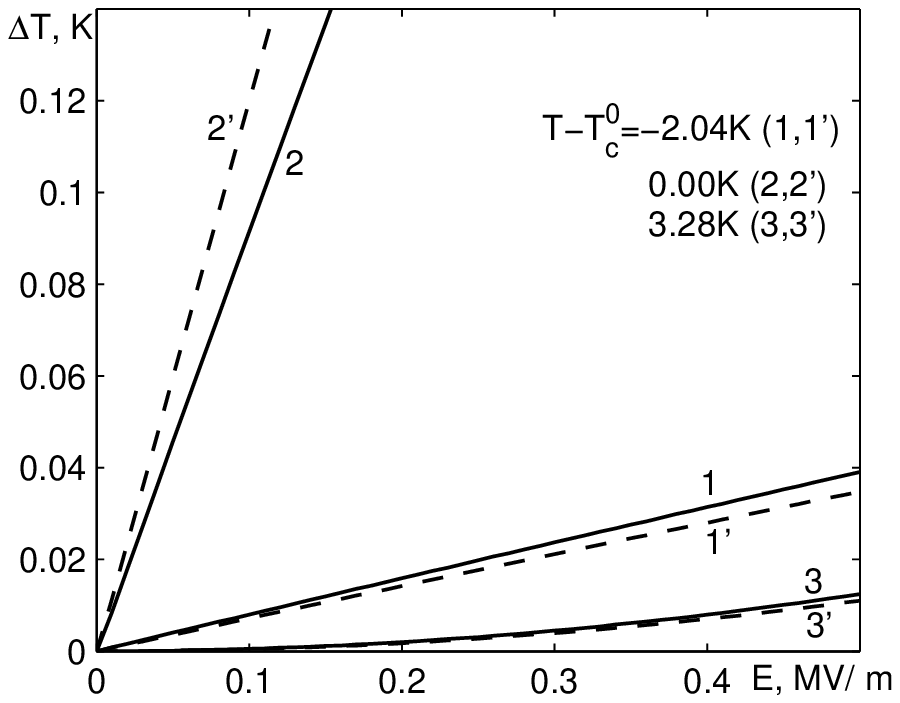}
 \hspace{4mm}
 \includegraphics[width=0.45\textwidth]{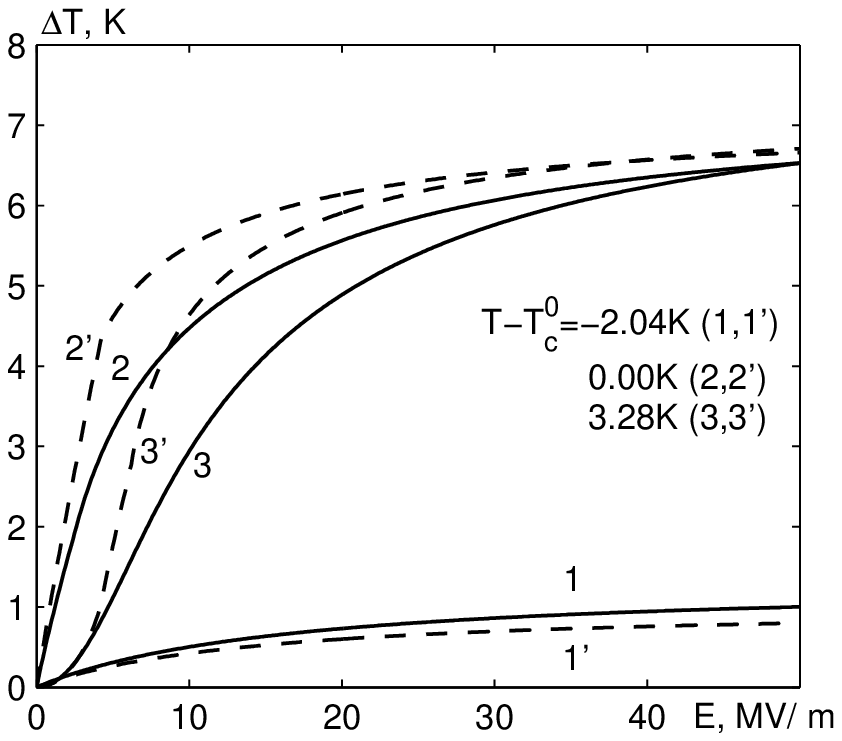}
\end{center}
\caption[]{The field dependence of the electrocaloric temperature
change of KH$_{2}$AsO$_{4}$  (solid lines) and  KD$_{2}$AsO$_{4}$
(dashed lines)  at $T-T_\textrm{c}^0=-2.04$~K~--- 1, $1'$; $T=T_\textrm{c}^0$~--- 2,
$2'$; $T-T_\textrm{c}^0=3.2$~K~--- 3, $3'$.} \label{DT_x89_KDA}
\end{figure}

The calculated electrocaloric changes of temperature $\Delta T$ of
the K(H$_{1 - x}$D$_{x})_{2}$PO$_{4}$ and K(H$_{1 -
x}$D$_{x})_{2}$AsO$_{4}$ crystals with the adiabatically applied
electric field  are shown in figures~\ref{DT_x89} and
\ref{DT_x89_KDA}. The experimental data of \cite{Shimshoni1969}
were obtained at $T=121$~K, which was very close to the transition
temperature of the sample used in the
measurements.%

As one can see, at small fields (figures~\ref{DT_x89},
\ref{DT_x89_KDA}, left-hand) the calculated electrocaloric temperature
change is a linear function of the field below $T_\textrm{c}^0$ (curves 1,
$1'$) and a quadratic function above $T_\textrm{c}^0$ (at 2, $2'$). The
experimental behavior below $T_\textrm{c}^0$ is not linear at $E_3<2$~kV/cm
due to the domains: The domains, whose polarization is
oriented along the field, are heated,
  whereas the domains, polarized in the opposite direction are cooled, thus the resulting net change of the sample temperature
  is close to zero.  The   experimental data for the
  electrocaloric temperature change at and above $T_\textrm{c}^0$ available for KH$_{2}$PO$_{4}$, as well as
 the $\Delta T/\Delta E$ ratio below $T_\textrm{c}^0$ at fields
above 2~kV/cm (when the sample is in a single-domain state), are
well reproduced by the theory.

At higher fields (figures~\ref{DT_x89}, \ref{DT_x89_KDA}, right-hand) the
calculated electrocaloric temperature changes at temperatures
above $T_\textrm{c}^0$ are larger than below $T_\textrm{c}^0$. The obtained curves
deviate from linear and quadratic behavior and reach saturation at
$E\gg500$~kV/cm. It should be mentioned, however, that these
curves are calculated with the linear over the field $E_3$
pseudospin Hamiltonian (\ref{H_s}). It would be very interesting
to compare our results at high fields with experiment, for
instance, to find out when non-linear contributions to the
Hamiltonian cannot be omitted any longer. Unfortunately, no
experimental data for $\Delta T$ in the fields above 1~kV/cm are
available. And, of course, possibilities for experimental
measurements are limited by the dielectric strength of the
samples.

\begin{figure}[!t]
\begin{center}
 \includegraphics[width=0.44\textwidth]{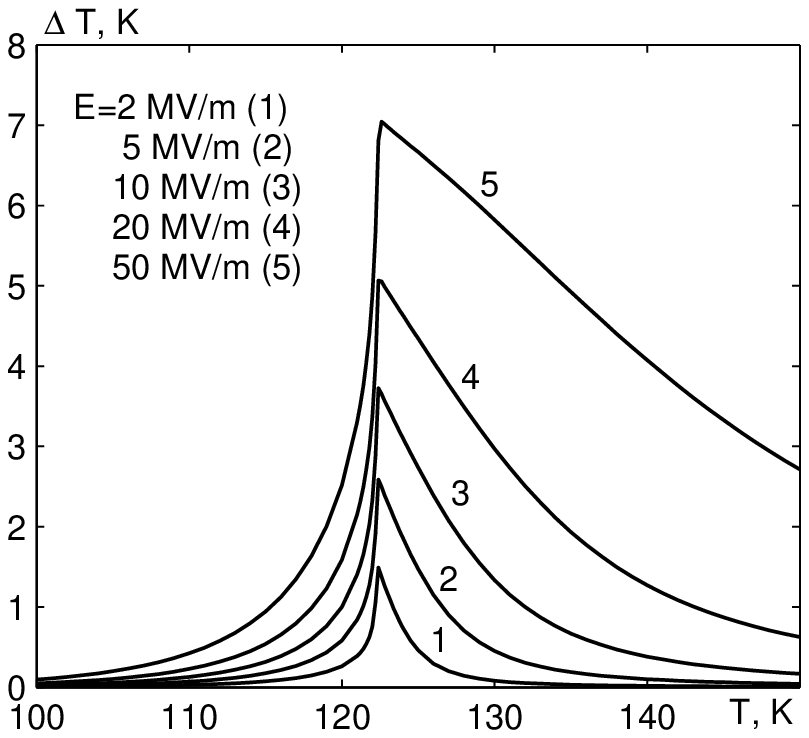}
 \hspace{5mm}
 \includegraphics[width=0.45\textwidth]{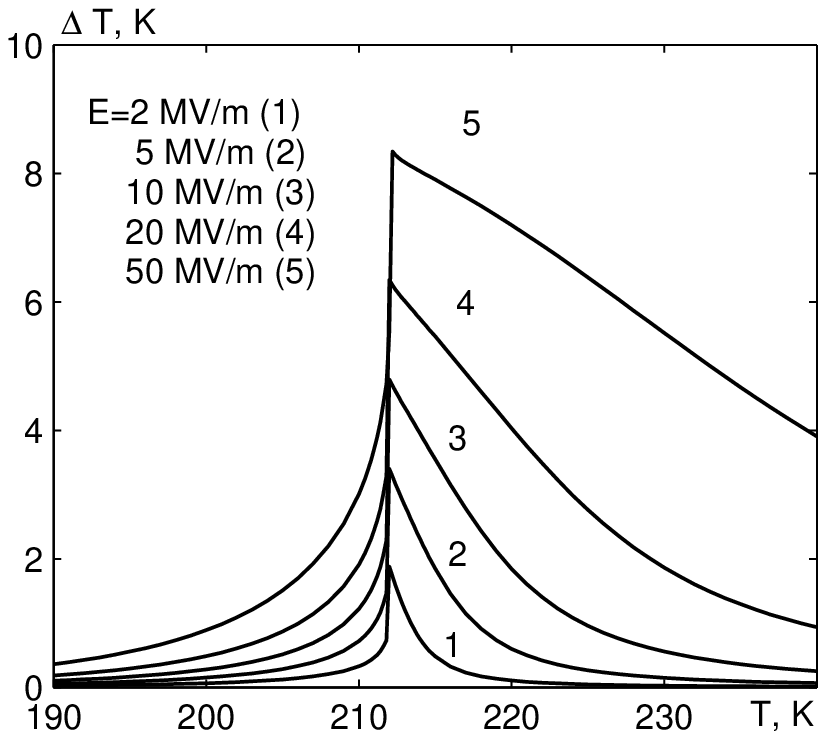}
\end{center}
\caption[]{The temperature dependence of the electrocaloric
temperature change of K(H$_{1 - x}$D$_{x})_{2}$PO$_{4}$ for
$x=0.0$ (left-hand) and $x=0.89$ (right-hand) in different fields.}
\label{DT_T}
\end{figure}

As one can see from the temperature dependence of  $\Delta T$
(figure~\ref{DT_T}) for K(H$_{1 - x}$D$_{x})_{2}$PO$_{4}$ crystals,
the calculated electrocaloric temperature change is the largest at
temperatures below $T_\textrm{c}^0$ but close $T_\textrm{c}$ and can exceed 6~K;
however, the fields required to reach $\Delta T$ that high are not
accessible in reality, because most likely they exceed the dielectric
strength of the crystals.

 \section{Conclusions}

Taking into account the dependence of the effective dipole moment
on the order parameter within the framework of the proton ordering
model allows us to correctly describe the smearing of the
ferroelectric phase transition in high electric fields as well as
the electrocaloric effect in the KDP family crystals. The theory
predicts the values of the electrocaloric temperature change of  a
few Kelvins in high fields.  Additional experimental measurements
of  $\Delta T$ in the fields above 2~kV/cm are necessary.

\appendix

\section*{Appendix}
The notations introduced in equations (\ref{csigma})--(\ref{csigma1})
are as follows:
 \bea
 N_6 &=& 2a(\beta\varepsilon)^2\cosh\beta\delta_{a}\varepsilon_6 +4
b(\beta
w)^2\cosh(z  -\beta\delta_{1}\varepsilon_6) +(\beta w_1)^2d
+2\varepsilon_6
\beta^2(-\varepsilon\delta_{a}M_{a}+w\delta_{1}M_{1})
\nonumber\\
&&  +\,\varepsilon_6^2
\left[2a(\beta\delta_{a})^2\cosh\beta\delta_{a}\varepsilon_6 +
(\beta\delta_{s})^2\cosh(2z  +\beta\delta_{s}\varepsilon_6) + 4
b(\beta\delta_{1})^2\cosh(z  -\beta\delta_{1}\varepsilon_6)\right],
\nonumber
\eea
\be
q_6 = 2b \beta w \sinh(z
-\beta\delta_{1}\varepsilon_6)
+\varepsilon_6\beta\left[-\delta_{s}\cosh(2z
+\beta\delta_{s}\varepsilon_6) + 2b\delta_{1}\cosh(z
-\beta\delta_{1}\varepsilon_6)\right],
 \nonumber
\ee
\be
\lambda =-\beta\varepsilon\delta_{a}M_{a}+\beta
w\delta_{1}M_{1}
+ \varepsilon_6
\beta\left[\delta_{s}^2\cosh(2z +\beta\delta_{s}\varepsilon_6) +
2a\delta_{a}^2\cosh\beta\delta_{a}\varepsilon_6 + 4b\delta_{1}^2
\cosh(z  -\beta\delta_{1}\varepsilon_6)\right],
 \nonumber
 \ee
\be
 \eta_T = p_6^\varepsilon + \frac{v}{2(\mu+12\mu'\eta^2)} (e_{36}-e_{36}^2)\varepsilon_T\,,   \nonumber
\ee
\be
\varepsilon_T = \left[\frac{2\beta}{vD}\left(2Tz_Tf_6-\lambda +\frac{M r }{D}\right) - \frac{4
 p_6^{\varepsilon}}{v}\left(\psi_6-\frac{z_{\eta}f_6}{D}\right)\right]\Big/c_{66}^{E}\,.
 \nonumber
 \ee
 In  turn,
\[
 p_6^{\varepsilon} = \frac1T\frac{2\varkappa Tz_T + [q_6 -\eta M ]}{D-2\varkappa
 z_{\eta}}\,,
\]
 $ c_{66}^{E}$ is
 the isothermal elastic
constant at a constant field
 \bea
 c_{66}^{E} &=& c_{66}^{E0} + \frac{8\psi_6}{v}
 \frac{\beta(- \psi_6 \varkappa  + f_6)}{D - 2 z_{\eta}
 \varkappa } -
 \frac{4 \beta z_{\eta} f_6^2}{v D (D - 2 z_{\eta}
 \varkappa )}\nonumber \\
 && - \frac{2\beta}{v D} \left[\delta_{s}^2 \cosh (2z  + \beta \delta_{s} \varepsilon_6) +
2a \delta_{a}^2 \cosh \beta\delta_{a}\varepsilon_6 + 4b
\delta_{1}^2 \cosh (z  - \beta \delta_{1} \varepsilon_6)
 \right] + \frac{2\beta r ^2}{v D^2}\,,\nonumber
 \eea
and $e_{36}$ is the isothermal piezoelectric coefficient
\bea && \hspace{-4ex} e_{36} = - \left(
\frac{\partial\sigma_6}{\partial E_3} \right)_{T,\varepsilon_6} =
\left( \frac{\partial P_3}{\partial\varepsilon_6}\right)_{T,E_3} =
e^0_{36} + \frac{2(\mu + 12\mu'\eta^2)}{v}
\frac{\beta\theta_6}{D-2 z_{\eta}\varkappa }\,,\nonumber
\eea
with
\bea &&
\theta_6 = - 2 \varkappa \psi_6 + f_6\,,\qquad  f_6=\delta_{s} \cosh
(2z  +\beta\delta_{s}\varepsilon_6) - 2b\delta_{1}\cosh(z
-\beta\delta_{1}\varepsilon_6)+\eta r ,\nonumber\\&&  z_{\eta} =
\frac{1}{1- \eta ^2} + \beta\nu_\textrm{c} + 12\beta\mu'\eta E_3\,.\nonumber
\eea

\clearpage

\ukrainianpart

\title{Електрокалоричний ефект у кристалах типу KH$_2$PO$_4$}
\author{А.С. Вдович\refaddr{label1}, А.П. Моїна\refaddr{label1}, Р.Р. Левицький\refaddr{label1}, І.Р. Зачек\refaddr{label2}}
\addresses{
\addr{label1} Інститут фізики конденсованих систем НАН України,  вул. І.~Свєнціцького, 1,
79011  Львів, Україна
\addr{label2} Національний університет ``Львівська політехніка'', вул. С.~Бандери, 12, 79013 Львів, Україна }

\sloppy

\makeukrtitle

\begin{abstract}
В моделі протонного впорядкування для кристалів типу KH$_{2}$PO$_{4}$ враховано
залежність  ефективних дипольних моментів від параметра протонного впорядкування.
В наближенні чотиричастинкового кластера  розраховано поляризацію кристалів та досліджено електрокалоричний ефект у них. Описано
розмивання сегнетоелектричного фазового переходу поздовжним електричним полем. Отримано добре узгодження з експериментальними даними.
\keywords електрокалоричний ефект, KDP, кластерне наближення,
поляризація
\end{abstract}

\end{document}